\documentclass[prd,twocolumn,showpacs,superscriptaddress,groupedaddress,nofootinbib]{revtex4}  
\usepackage{graphicx}  
\usepackage{lipsum}
\usepackage{dcolumn}   
\usepackage{bm}        
\usepackage{amsmath,amssymb}   
\usepackage{aas_macros}
\usepackage{multirow}
\usepackage{color}
\usepackage{verbatim}
\usepackage{url}
\usepackage{hyperref}
\usepackage{csquotes}
\usepackage{natbib}

\newcommand{\bs}{\boldsymbol}

\newcommand{\Largr}{\mathcal{L}}
\newcommand{\cites}[1]{\citeauthor{#1}'s}

\begin{document}
\widetext
\title{Increasing Fisher Information by Moving-Mesh Reconstruction}
\author{Qiaoyin Pan}
\email{panda@mail.nankai.edu.cn}
\affiliation{School of Physics, Nankai University, 94 Weijin Rd, Nankai, Tianjin, 300071, China}
\affiliation{Canadian Institute for Theoretical Astrophysics, University of Toronto, 60 St. George Street, Toronto, Ontario M5S 3H8, Canada}
\author{Ue-Li Pen}
\email{pen@cita.utoronto.ca}
\affiliation{Canadian Institute for Theoretical Astrophysics, University of Toronto, 60 St. George Street, Toronto, Ontario M5S 3H8, Canada}
\affiliation{Dunlap Institute for Astronomy and Astrophysics, University of Toronto, Toronto, ON M5S 3H4, Canada}
\affiliation{Canadian Institute for Advanced Research, Program in Cosmology and Gravitation} 
\affiliation{Perimeter Institute for Theoretical Physics, Waterloo, ON, N2L 2Y5, Canada}
\author{Derek Inman}
\affiliation{Canadian Institute for Theoretical Astrophysics, University of Toronto, 60 St. George Street, Toronto, Ontario M5S 3H8, Canada}
\affiliation{Department of Physics, University of Toronto, 60 St. George, Toronto, ON M5S 1A7, Canada}
\author{Hao-Ran Yu}
\affiliation{Canadian Institute for Theoretical Astrophysics, University of Toronto, 60 St. George Street, Toronto, Ontario M5S 3H8, Canada}
\affiliation{Kavli Institute for Astronomy and Astrophysics, Peking University, Beijing 100871, China}

\date{\today}

\begin{abstract}
  Reconstruction techniques are commonly used in cosmology to reduce
  complicated nonlinear behaviours to a more tractable linearized
  system.  We study a new reconstruction technique that uses the
  Moving-Mesh algorithm to estimate the displacement field
  from nonlinear matter distribution. We show the performance of this
  new technique by quantifying its ability to reconstruct linear
  modes. We study the cumulative Fisher information $I(<k_n)$ about  
  the initial matter power spectrum in the matter power spectra in 130 $N$-body simulations before 
  and after reconstruction, and find that the nonlinear plateau of
  $I(<k_n)$ is increased by a factor of $\sim 50$ after reconstruction, from
  $I \simeq 2.5 \times 10^{-5} /({\rm Mpc}/h)^3$ to
  $I \simeq 1.3 \times 10^{-3}/({\rm Mpc}/h)^3$ at large $k$.
  This result includes the decorrelation between initial and final fields,
  which has been neglected in some previous studies.
We expect this technique to be
  beneficial to problems such as baryonic acoustic oscillations,
  redshift
space distortions and
  cosmic neutrinos that rely on accurately disentangling nonlinear
  evolution from underlying linear effects.
\end{abstract}

\maketitle

\begin{section}{Introduction}\label{sec:introduction}  

  Two-point statistics provide complete descriptions of Gaussian
  density fields and can be computed efficiently even for large data
  sets.  However, nonlinear gravitational evolution leads to highly
  non-Gaussian matter distributions which require higher order
  statistics to fully characterize.  Such statistics are
  computationally expensive and can be challenging to map to
  cosmological parameters.  To mitigate these difficulties, it is
  common to transform the matter field in a way that hopefully reduces
  non-Gaussianity.  For example, Gaussianization transforms have been
  used to make the logarithmic distribution more Gaussian
  \cite{bib:Weinberg1992,bib:Mark2009} and wavelet nonlinear Wiener
  filters have been used to separate Gaussian and non-Gaussian
  components of the density field
  \cite{bib:Zhang2011,bib:Yu2012,bib:HarnoisD2013}.
  {\it Reconstruction} techniques \citep{bib:Daniel2007} provide a more
  effective way by converting matter distributions back to an
  earlier stage \citep{bib:HarnoisD2013}.

  The quality of these techniques can be quantified by computing the Fisher
  information \citep{bib:Rimes2006} present in the power spectrum before and after
  reconstruction/Gaussianization.
  \citet{bib:Martin1999} were the first to study the Fisher information in the nonlinear
  matter power spectrum calculated from $N$-body simulations.  They
  found that the cumulative information has a plateau on translinear scales
  ($k \simeq 0.2-0.8$ $h$/Mpc) due to strong coupling between Fourier
  modes.  Qualitatively, this means that the power spectra on these
  scales give little additional information.  
  \citet{bib:Ngan2012} applied the linear reconstruction using the Zel'dovich approximation 
  on nonlinear density fields, and found that the cumulative Fisher information increases slightly.
  \citet{bib:HarnoisD2013} computed the cumulative Fisher information
  for various Gaussianization methods and their combinations
  and found that while mode coupling is reduced, there is not
  necessarily an improvement in the cross correlation between the
  initial density fields and the final nonlinear ones. 

  In studies of Baryon Acoustic Oscillations (BAO), density fields are
  subjected to reconstruction which partially inverts nonlinear
  evolution by applying the negative Zel'dovich displacement field \cite{bib:Eisenstein2007,bib:Zel1970}.
  \citet{bib:Yu2016} studied the curl-free, or $E$-mode, component of 
  the exact displacement field in Lagrangian
  space and found that the linear density field can be well recovered by
  the $E$-mode displacement field.
  This $E$-mode reconstruction therefore provides a theoretical 
  target for other reconstruction techniques to be compared with.
  Recently \citet{bib:Zhu2016,bib:ZhuH2016} described a new reconstruction 
  technique using the Moving-Mesh
  algorithm (MM), first described in \cite{bib:Pen1995,bib:Pen1998},
  to effectively estimate $\bs{\Psi}(\bs{q})$ from only nonlinear
  density fields.  They further showed that even though shell-crossing
  and vorticity are not recovered, linear density modes are still 
  recovered up to scales relevent to the BAO.

  In this paper, we compute the Fisher information recovered after
  using this new reconstruction scheme on 130 independent $N$-body
  simulations, and compared with other methods and unreconstructed
  fields.  The paper is organized as follows.  In \S
  \ref{sec:reconstruction}, we briefly
  describe the computation of the displacement potential using MM
  algorithm. In \S \ref{sec:simulation}, we describe  
  the simulations, implementation of the reconstruction and compare 
  the power spectra and cross correlations before and after reconstruction.  
  In \S \ref{sec:fisherinfo}, we further compute the correlation matrix 
  and Fisher information before and
  after reconstruction.  Finally, in \S \ref{sec:conclusion}, we summarize 
  our results and 
  discuss the effectiveness of the reconstruction.

\end{section}

\begin{section}{Reconstruction Technique}
  \label{sec:reconstruction}
  Here we briefly review the MM algorithm used in the new reconstruction technique; for a more
  complete description we refer the reader to \cite{bib:ZhuH2016}.  
  The aim of the MM algorithm is to estimate the displacement field of mass elements in 
  Lagrangian coordinates from their final Eulerian position only, and from this
  estimated displacement field one directly reconstructs the linear density field. The
  general principle is to relate the Eulerian coordinates of a mass element, $x^i$ to
  a curvilinear system, $\xi^\mu$, such that the mass
  per grid cell is approximately constant,
  \begin{align}
   \label{eq:const}
    \rho \sqrt{g}=\mathrm{Const.},
  \end{align}
  where $\sqrt{g} \equiv \mathrm{det}\left| e^i_\mu\right|$ is the volume
  element and $e^i_\mu \equiv \partial x^i / \partial \xi ^ \mu$ is the coordinate transformation matrix. 
  These coordinates are
  related via a {\it deformation} field, which we assume to be a pure
  gradient:
  \begin{align}
    x^i = \xi^\mu \delta^i_\mu + \frac{\partial \phi}{\partial
    \xi^\mu}\delta^{i\mu},
  \end{align}
  and $\phi$ is called the {\it deformation potential} which is chosen to satisfy Eq. \ref{eq:const}.  

  Numerically, we iteratively solve for
  the deformation potential via a diffusion equation, 
  \begin{align}
    \label{eq:li_elip}
    \partial _\mu (\rho \sqrt{g} e^\mu _i \delta^{i\nu}
    \partial_\nu \dot{\phi})=\Delta \rho,
  \end{align}
  where $\Delta \rho = \langle\rho\rangle-\rho \sqrt{g}$ is the difference in density 
  due to displacing the grids. A detailed description 
  of the analytical formulation can be found in the adaptive
  particle mesh and moving mesh (MM) hydrodynamics algorithms \cite{bib:Pen1995,bib:Pen1998}.
  Eq.~\ref{eq:li_elip} can be solved by multi-grid
  algorithm\cite{bib:Pen1995,bib:Pen1998,bib:ZhuH2016}.
  Then the estimated displacement field is given by
  \begin{align}
   \label{eq:disp}
   \bs{\tilde{\Psi}}(\bs{\xi})=\nabla \phi(\bs{\xi}),
  \end{align}
  and the reconstructed density field is given by
  \begin{align}
   \label{eq:delta}
   \delta_R(\bs{\xi})=- \nabla \cdot \bs{\tilde{\Psi}}(\bs{\xi}) = - \nabla^2 \phi(\bs{\xi}). 
  \end{align}

\end{section}

\begin{section}{Implementation and Power Spectra}
  \label{sec:simulation}
  We use \textsc{CUBEP$^3$M} \cite{bib:Harnois2013} to run
  140 simulations with a box size of 600 Mpc/$h$ and $512^3$ particles.
  For these simulations, we use cosmological
  parameters $\Omega_m=0.321$, $\Omega_{\Lambda}=1-\Omega_m=0.679$,
  $h=0.67$, $\sigma_8=0.83$, and $n_s=0.96$.
  The initial conditions are computed by
  transfer function \cite{bib:Lewis2000}
  at $z=100$.
  Zel'dovich
  approximation is used to calculate the displacement and initial velocities
  of particles.

  \begin{figure*}[t!]
    \centering
    \includegraphics[width=0.9\textwidth]{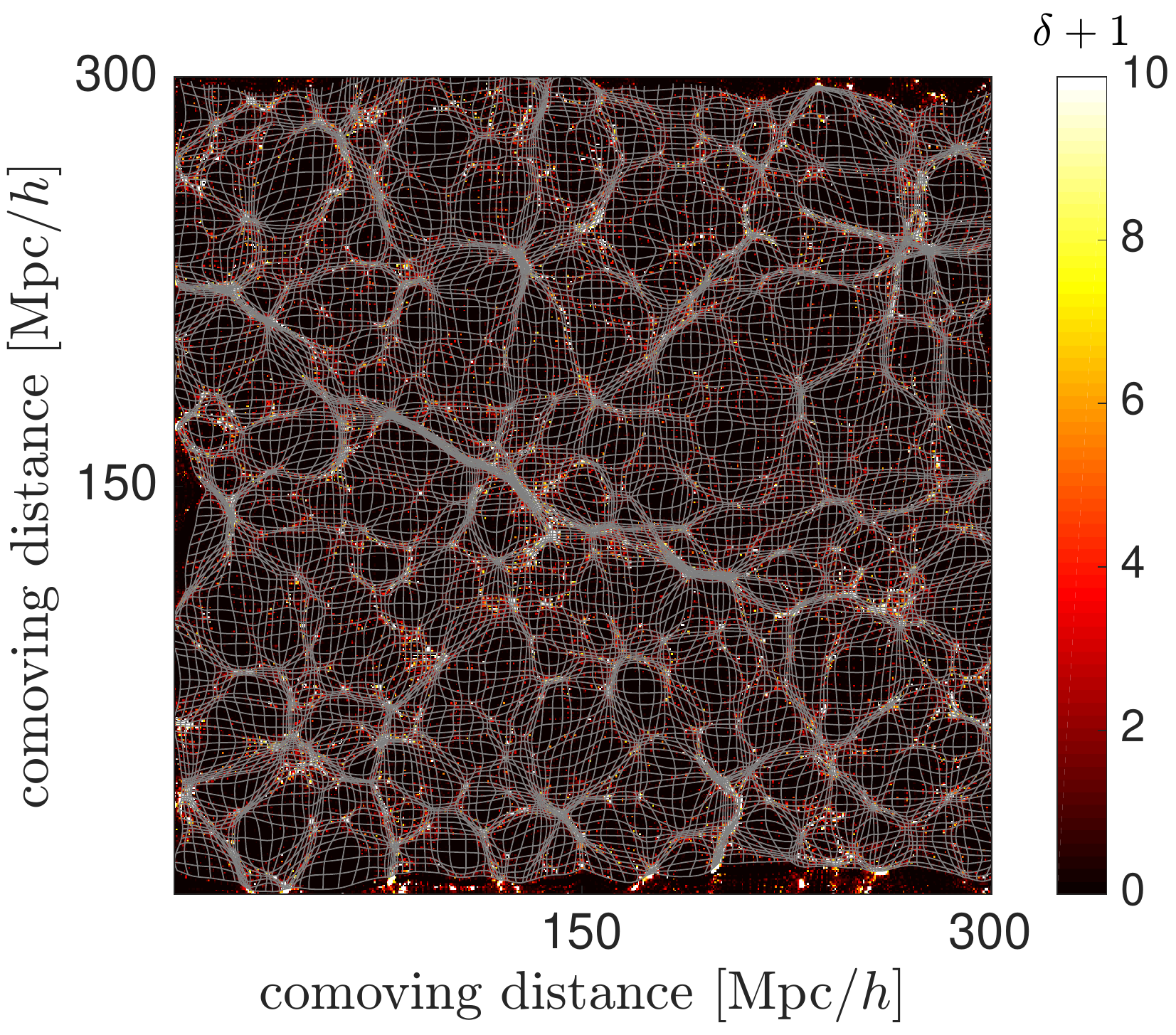}
    \caption{ Illustration of MM reconstruction.
      The 2-D projection of one layer of the deformed mesh of a sample
      $N$-body simulation is shown as curved white lines.  The
      density $\rho/\langle\rho\rangle=1+\delta$ on the mesh is shown
      underneath. For clarity, the scale of the density field is cut to 
      300 Mpc/$h$, and only every other grid line is plotted.}
    \label{fig:simandrec}
 \end{figure*}

 We use the Voronoi tessellation method \cite{bib:Van1994} to estimate the density contrast
 $\delta_N=\rho/\langle\rho\rangle-1$ ($N$ stands for nonlinear) from particle distributions, and apply the
 MM reconstruction to these fields with $512^3$ grids.
 The reconstruction code solves the displacement potentials iteratively
 until the root mean square (rms) of the results drop from $\sim 7.5$
 to 0.20. The compression limiter is set to be 0.1 \cite{bib:Pen1995, bib:Pen1998,bib:ZhuH2016}. 
 For different simulation samples, a different number of
 iterations are required to get the results of the same rms. A total of
 130 simulations converged to the target rms within 2000 iterations, 
 and we use these results for the calculation in this paper.
 A 2-D projection
 of one layer of the deformed mesh and the original density field on
 the mesh are shown in Fig.~\ref{fig:simandrec}. 
 As expected, the deformed mesh traces the structure very well and
 the deformed grids do not cross each other, even in the 2-D projection.
 
 We study the Fisher information in the matter power
 spectra. More generally,
 the cross power spectrum $P_{\alpha\beta}(k)$ for species $\alpha$ and $\beta$
 ($\alpha=\beta$ for auto power spectrum) is defined as
 \begin{align}
   \langle \delta_\alpha^\dagger(\bm{k})\delta_\beta(\bm{k}') \rangle =
   (2\pi)^3 P_{\alpha\beta}(k) \delta_{\rm {3D}}(\bm{k}-\bm{k}'),
 \end{align}
 where $\delta_{\alpha}$ and $\delta_{\beta}$ are any two fields and
 $\delta_{\rm{3D}}$ is the three-dimensional Dirac delta function. We typically consider instead
 the dimensionless power spectrum, $\Delta_{\alpha\beta}^2(k)$, defined as
 \begin{align}
   \Delta_{\alpha\beta}^2(k) \equiv \frac{k^3 P_{\alpha\beta}(k)}{2\pi ^2}.
 \end{align}
 In the left panel of Fig.~\ref{fig:cp}, we show the matter auto power
 spectrum of linear theory density fields $\delta_L$, nonlinear density
 fields ($\delta_N$) from simulations and reconstructed density fields
 (from Eq. \ref{eq:delta}).  For the simulation
 results, we use the average value of all 130 simulations and show
 $1\sigma$ standard deviations as error bars.  

 To quantify the cross-correlation
 between fields, we compute the cross correlation coefficient
 $r_{\alpha\beta}(k)\equiv P_{\alpha\beta}/\sqrt{P_{\alpha\alpha}P_{\beta\beta}}$.  In the right panel of
 Fig.~\ref{fig:cp}, we show $r_{NL}$ and $r_{RL}$.  We see that, compared with $\delta_N$,
 $\delta_R$ correlates with $\delta_L$ on much wider range of scales.
 We compare our reconstruction correlation coefficient to that of the $E$-mode 
 reconstruction, $r_{EL}$, 
 computed in \citet{bib:Yu2016}.
 Even though $r_{RL}$ decreases from $r_{EL}$ in the nonlinear regime due to the fact that the MM reconstruction 
 cannot recover the shell-crossing present on these scales, we find that linear
 modes are recovered successfully on scales $k\simeq 0.05 - 0.3$ $h$/Mpc.
 Specifically, the scale where $r(k)=1/2$ increases from $k\simeq 0.2$ $h$/Mpc to
 $0.8$ $h$/Mpc after reconstruction.  In comparison with the results of \citet{bib:ZhuH2016},
 which showed $r(k\simeq0.9 h/\rm{Mpc})=1/2$, we find the correlation coefficient falls off at slightly lower
 wave numbers, which we attribute to using fewer particles per simulation.

  \begin{figure*}
    \centering
    \includegraphics[width=0.5\textwidth]{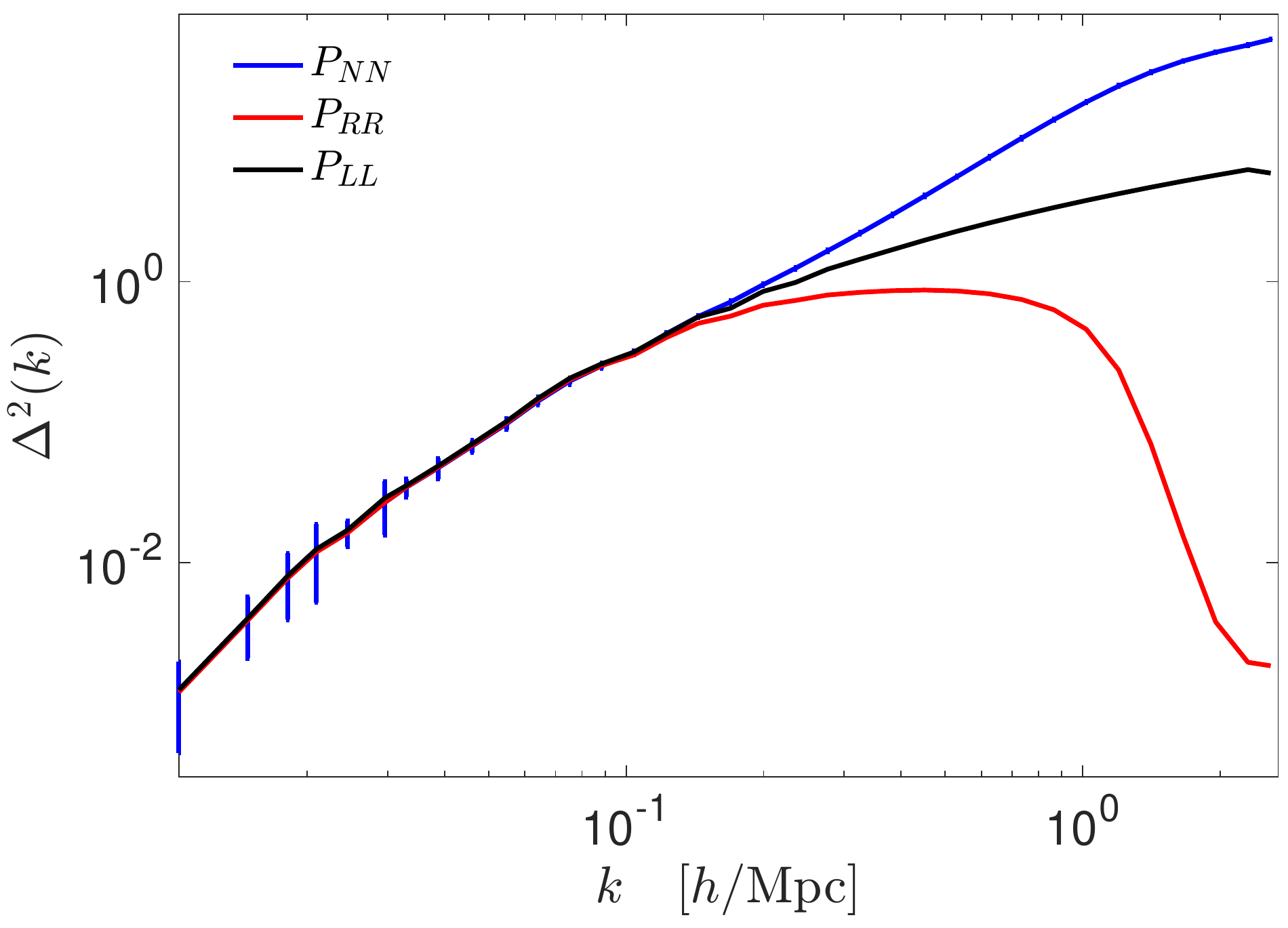}
    \includegraphics[width=0.485\textwidth]{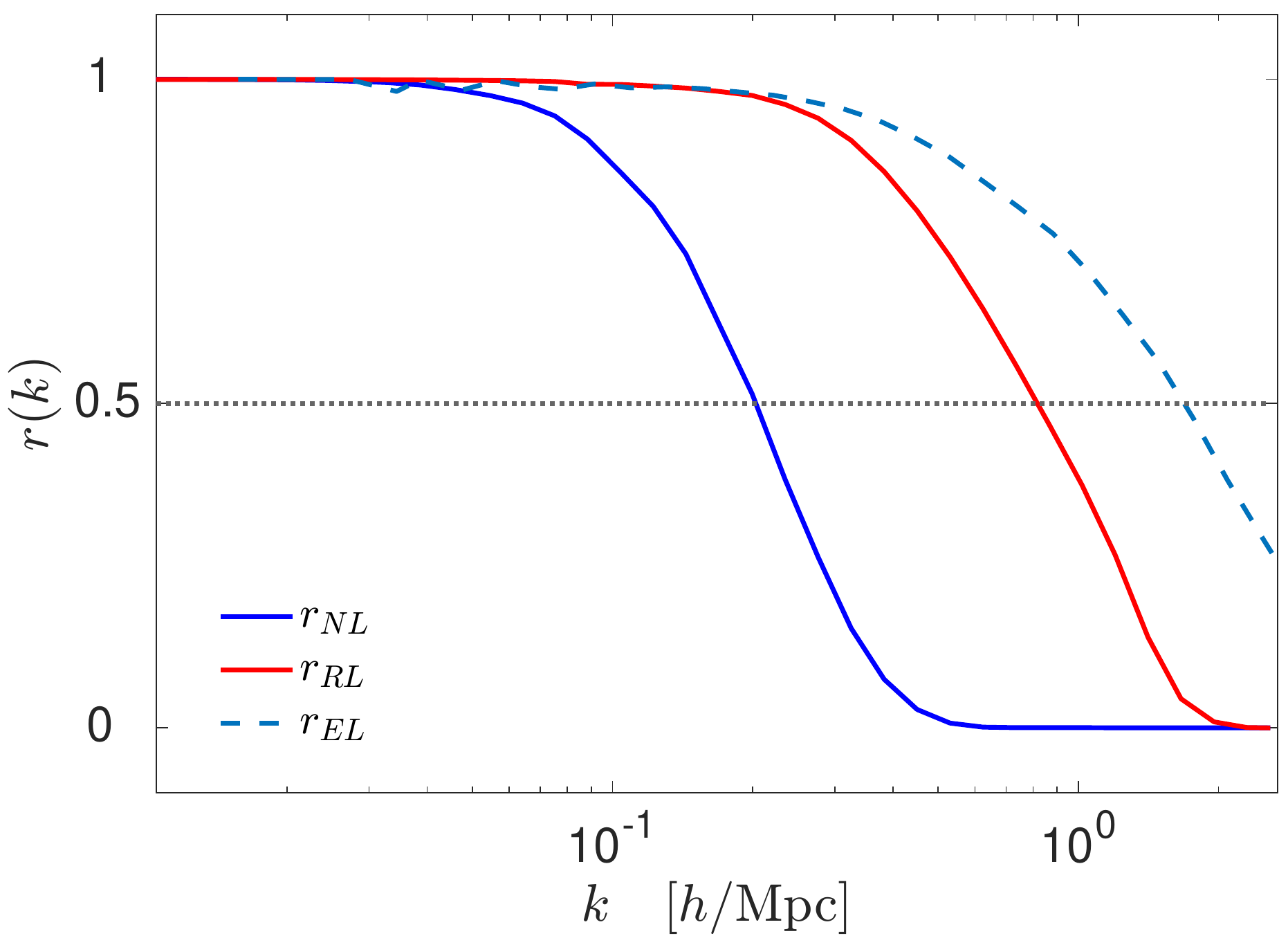}
    \caption{{\it Left.} The dimensionless power spectrum computed via
      linear theory (black), the mean value of 130 $N$-body
      simulations with $1\sigma$ error bars (blue), and reconstruction
      of the simulations (red).  {\it Right.} The cross correlation
      coefficient between simulation and linear densities $r_{NL}$ (blue),
      MM reconstructed and linear densities $r_{RL}$ (red), and E-mode reconstruction $r_{EL}$ (dashed
      blue) from \citet{bib:Yu2016}.}
    \label{fig:cp}
  \end{figure*}

\end{section}

\begin{section}{Fisher Information Content}
  \label{sec:fisherinfo}
  Mathamatically, the Fisher information $I$ of the initial scale
  invariant matter power spectrum, $A$, is defined as
  \begin{align}
    I_A \equiv -\left\langle \frac{\partial ^2 \mathrm{ln \Largr}}{\partial  \mathrm{ln} A ^2}\right\rangle,
    \label{eq:fisherdefine}
  \end{align}
  where $\Largr$ is the likelihood \cite{bib:Tegmark1997}.  
  In this paper, the word \enquote{information} and the symbol \enquote{$I$} both implicitly 
  mean cumulative Fisher information 
  of $A$. For Gaussian
  fields, the likelihood depends on parameters only through the
  power spectrum $P(k)$, so $I$ can be written as 
  \begin{align}
    I = - \left\langle \sum_{k,k'} \frac{\partial \mathrm{ln} P(k)}{\partial \mathrm{ln} A} 
    \frac{\partial ^2 \mathrm{ln \Largr}}{\partial \mathrm{ln} P(k) \partial \mathrm{ln} P(k')}
    \frac{\partial \mathrm{ln} P(k')}{\partial \mathrm{ln} A}\right\rangle,
    \label{eq:fisherdef2}
  \end{align}
  where the angle bracket averages over realizations
  \cite{bib:Rimes2006}.
  Eq.~\ref{eq:fisherdef2} can be written in a simpler
  form in two aspects.   
  
  Firstly, we simplify the derivative term
  $\partial \mathrm{ln} P(k)/\partial\mathrm{ln} A$.  For a given density field $\delta_\alpha$, we can
  conveniently decompose it into a correlated, linear component,
  and an uncorrelated, noise component with respect to $\delta_L$, 
  \begin{align}
    \delta_{\alpha}(k) = r'(k) \delta_L (k) + \delta_{n}(k),
    \label{eq:decompose}
  \end{align}
  where $\delta_{n}(k)$ is defined such that the correlation
  $\langle \delta_L^\dagger(k)\delta_{n}(k) \rangle=0$.  To solve $r'$, we correlate
  both sides with $\delta_L$ and the uncorrelated noise term drops out,
  \begin{align}
    \langle \delta_L^\dagger(k)\delta_\alpha(k) \rangle = r'(k) \langle \delta_L^\dagger(k)\delta_L(k) \rangle.
    \label{eq:correlating}
  \end{align} 
  Using the definitions of cross correlation coefficient, $r_{\alpha L}(k)\equiv P_{\alpha L}/\sqrt{P_{\alpha\alpha}P_{LL}}$ 
  and bias, $b^2(k)\equiv P_{\alpha\alpha}/P_{LL}$, we can solve for $r'$ as
  \begin{align}
    r'(k) = \frac{P_{\alpha L}(k)}{P_{LL}(k)}=r_{\alpha L}(k) b(k).
    \label{eq:bofk}                                              
  \end{align}                                                    
  To find the nonlinear term, we square both sides of Eq.~\ref{eq:decompose}
  and the cross term of the right hand side vanishes,             
  \begin{align}                                                  
    \langle \delta_\alpha^\dagger(k) \delta_\alpha(k) \rangle =  
    r_{\alpha L}^2(k)b^2(k) \langle \delta_L^\dagger(k) \delta_L(k) \rangle + 
    \langle \delta_{n}^\dagger(k)\delta_{n}(k) \rangle,          
  \end{align}                                                    
  and find                                                       
  \begin{align}                                                  
    P_{\alpha\alpha}(k) = r_{\alpha L}^2(k)b^2(k)P_{LL}(k) + P_{nn}(k).
    \label{eq:powerdecompose}                                    
  \end{align}                                                    
  With the help of Eq.~\ref{eq:bofk} and Eq.~\ref{eq:powerdecompose},
  we get                                                         
  \begin{align}                                                  
    \frac{\partial \mathrm{ln} P(k) }{ \partial \mathrm{ln} A}=
    r^2_{\alpha L}(k)b^2(k)\frac{P_{LL}(k)}{P_{\alpha\alpha}(k)}=r^2_{\alpha L}(k).
  \end{align}

  Secondly, we simplify
  $\partial ^2 \mathrm{ln \Largr}/\partial \mathrm{ln} P(k) \partial
  \mathrm{ln} P(k')$
  by using the fact that its expectation value is the Fisher
  matrix.  For Gaussian fields, this is equal to the inverse of the
  covariance matrix which is diagonal with elements given by the
  number of modes in each bin (when considering $\bs{k}$ and $-\bs{k}$ as the same mode).  
  We can extend this definition to
  non-Gaussian fields, by taking into account that the covariance
  matrix is no longer diagonal \cite{bib:Rimes2006}.  Thus, we
  write the Fisher information in terms of matrix multiplication:
  \begin{align}
    I \left( < k_n\right) = r^2(k)^{\mathrm{T}} \left[ \mathrm{C^{-1}_{norm}} 
    ( k,k' )\right]_{<k_n} r^2(k') ,
    \label{eq:fisherformulaused}
  \end{align}
  where
  \begin{align}
    \mathrm{C_{norm}} \left( k,k' \right)=\frac{\mathrm{Cov}(k,k')}
    {\langle P(k)\rangle\langle P(k')\rangle}
  \end{align}
  is the normalized covariance matrix, and
  $r$ is the mean cross correlation of a given density field with
  $\delta_L$ and the subscript $<k_n$ indicates the matrix elements are set to
  zero for modes $k,k'>k_n$.  The covariance matrix is defined as
  \begin{align}
    \mathrm{Cov}\left(k,k'\right)\equiv \frac{\sum_{i,j=1}^{N}\left[ P_i \left( k \right) - 
    \langle P \left( k \right) \rangle \right]\left[ P_j \left( k' \right) - 
    \langle P \left( k' \right)\rangle \right]}{N-1},
  \end{align}
  where $N$ is the total number of simulations and angle bracket average these simulations.  

  The cross-correlation coefficient matrix, or for short the correlation matrix, 
  is defined as 
  \begin{align}
    \mathrm{Corr}\left(k,k'\right)=\frac{\mathrm{Cov}\left(k,k'\right)}
    {\sqrt{\mathrm{Cov}\left(k,k\right)\mathrm{Cov}\left(k',k'\right)}},
  \end{align}
  representing the correlation between different $k$ modes.  The
  correlation matrices for nonlinear and reconstructed power spectra
  are shown in the upper-left and lower-right sections of Fig.~\ref{fig:corrall}.
  By definition, the correlation matrix is symmetric with unit
  diagonal allowing us to overlay the two matrices.  For the
  nonlinear case, it is almost diagonal in the linear
  regime, $k \lesssim 0.07$ $h$/Mpc.  The off-diagonal
  elements are produced by strong mode coupling on nonlinear scales
  and the super-survey tidal effect which is small on linear scales
  but dominates in the weakly nonlinear regime
  \cite{bib:Kazuyuki2016}.  The correlation matrix for the nonlinear
  power spectra has a small amount of negative elements
  ($\mathrm{Corr} \gtrsim -0.18$), which should vanish with more
  simulations \cite{bib:Takahashi2009}.  For the reconstructed
  correlation matrix, the linear regime extends up to $k \simeq 0.3$
  $h$/Mpc.  However, the number and magnitude of negative off-diagonal elements
  also increases ($\mathrm{Corr} \gtrsim -0.49$).

  \begin{figure}
    \centering
    \includegraphics[width=0.48\textwidth]{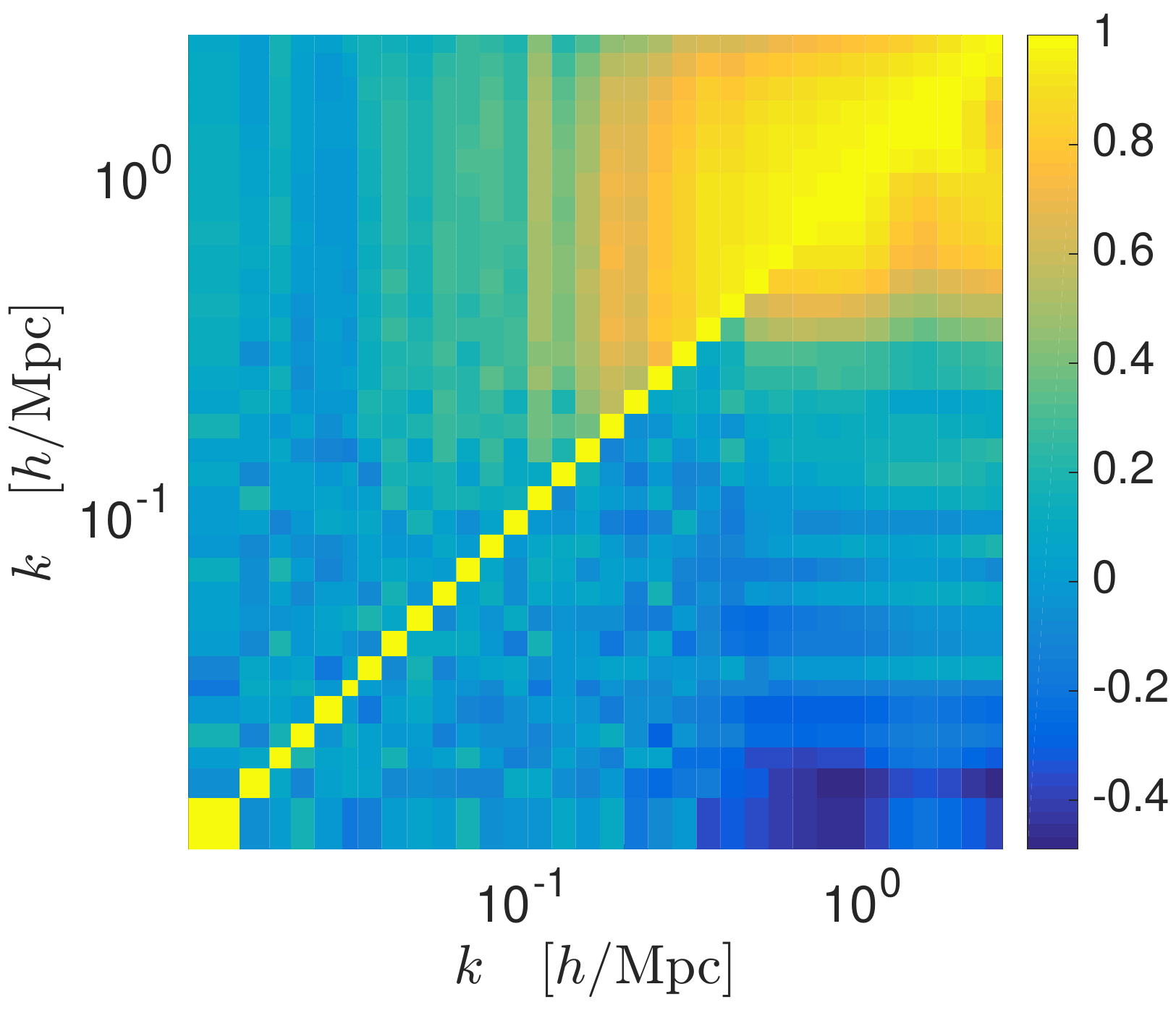}
    \caption{The correlation matrix from 130 nonlinear power
      spectra (the upper triangular elements) and reconstructed power
      spectra (the lower triangular elements).}
    \label{fig:corrall}
  \end{figure}

  The Fisher information is proportional to the volume. 
  We plot the Fisher information per unit volume of the power spectra of
  $\delta_N$, $\delta_L$ and $\delta_R$ in the left panel of 
  Fig.~\ref{fig:fisherinfo}. The Fisher information of the linear 
  power spectra is equal to the cumulative number of $k$-modes, $N_k$. As expected, Fisher information of
  $\delta_N$ decreases from that of $\delta_L$ on scale
  $k \simeq 0.05$ $h$/Mpc, and has a flat plateau on small scales, 
  with a saturated value of
  $I \simeq 2.5 \times 10^{-5}/({\rm Mpc}/h)^3$, indicating
  the absence of independent information in the nonlinear
  regime.  In comparison, the information curve of the $\delta_R$ power
  spectra keeps increasing roughly the same as the linear information
  until $k\simeq 0.3$ $h$/Mpc, and reaches a value of 
  $I \simeq 1.3 \times 10^{-3}/({\rm Mpc}/h)^3$ at $k \simeq 2.7$ $h$/Mpc,
  a factor of 50 times greater than $I_{\delta_N}$.
  As an example to illustrate its strength, we compare the Fisher information given by the MM reconstruction method
  with the logarithmic density mapping method \cite{bib:Mark2009}. We find that MM
  reconstruction gives over 10 times more information.  It also appears to perform 
  significantly better than the standard BAO reconstruction using the Zel'dovich 
  approximation \cite{bib:Ngan2012}, although a direct comparison is left for future work.
  
  To test the upper limit of information that the MM reconstruction can recover, 
  we also checked the Fisher information given by $\delta_E$ \cite{bib:Yu2016} as a reference, 
  which indicates the maximum possible information available via reconstruction procedures.
  We find that $I_{\delta_E}$ is 3 times higher ($150 I_{\delta_N}$) than MM reconstruction.
  This gives motivation to continue to develop and optimize our 
  reconstruction algorithms to achieve this theoretical target.
  
  In some previous works, the cross correlation $r^2$ terms have been set
  to unity in Eq.~\ref{eq:fisherformulaused}, which artificially
  increases the information. To demonstrate this, we plot this case in the right panel of 
  Fig.~\ref{fig:fisherinfo}.  We see that neglecting $r^2$ causes there to be an artificial 
  increase in information on small scales ($k\gtrsim 1h/\mathrm{Mpc}$).  We also see that 
  the information of the logarithmic density
  mapping is much higher than it is in the left panel.  Including the correlation coefficient 
  ensures that the information we compute is that present in the initial conditions and not 
  spuriously induced by the transformation procedure.  This therefore resolves the problems 
  discussed in \cites{bib:HarnoisD2013} \cite{bib:HarnoisD2013} section ``Information about what?''
  We conclude that, in the context of BAO analysis and extracting other primordial cosmological
  parameters, we should take into account the correlation term $r^2$ and
  use Eq.~\ref{eq:fisherformulaused} to compute the Fisher information.
  
  Another way to quantify the nonlinear scale of $\delta_\alpha$
  is via the information plateau's linear equivalent scale, $k_p$, satisfying
  \begin{equation}
      I_{\delta_L}(k_p)=I_{\delta_{\alpha}}(k\rightarrow\infty).
  \end{equation}
  In the left panel of Fig. \ref{fig:fisherinfo}, we can see that 
  $k_p$ is just the scale on which 
  the horizontal dotted line crosses $I_{\delta_{L}}$ curve.
  Practically, we use $I_{\delta_{\alpha}}$ ($k=2.7$ $h$/Mpc)
  as a proxy of $I_{\delta_{\alpha}}(k\rightarrow\infty)$,
  where $k$ is large enough such that $r\rightarrow 0$, ensuring
  the convergence, or saturation of $I_{\delta_{\alpha}}$.
  We find that
  for $\delta_N$, $k_p\simeq 0.15$ $h$/Mpc.
  The MM reconstruction increases $k_p$ to $0.4$ $h$/Mpc,
  whereas the logarithmic density mapping method only increases it to $0.19$ $h$/Mpc.

  \begin{figure*}
    \includegraphics[width=0.48\textwidth]{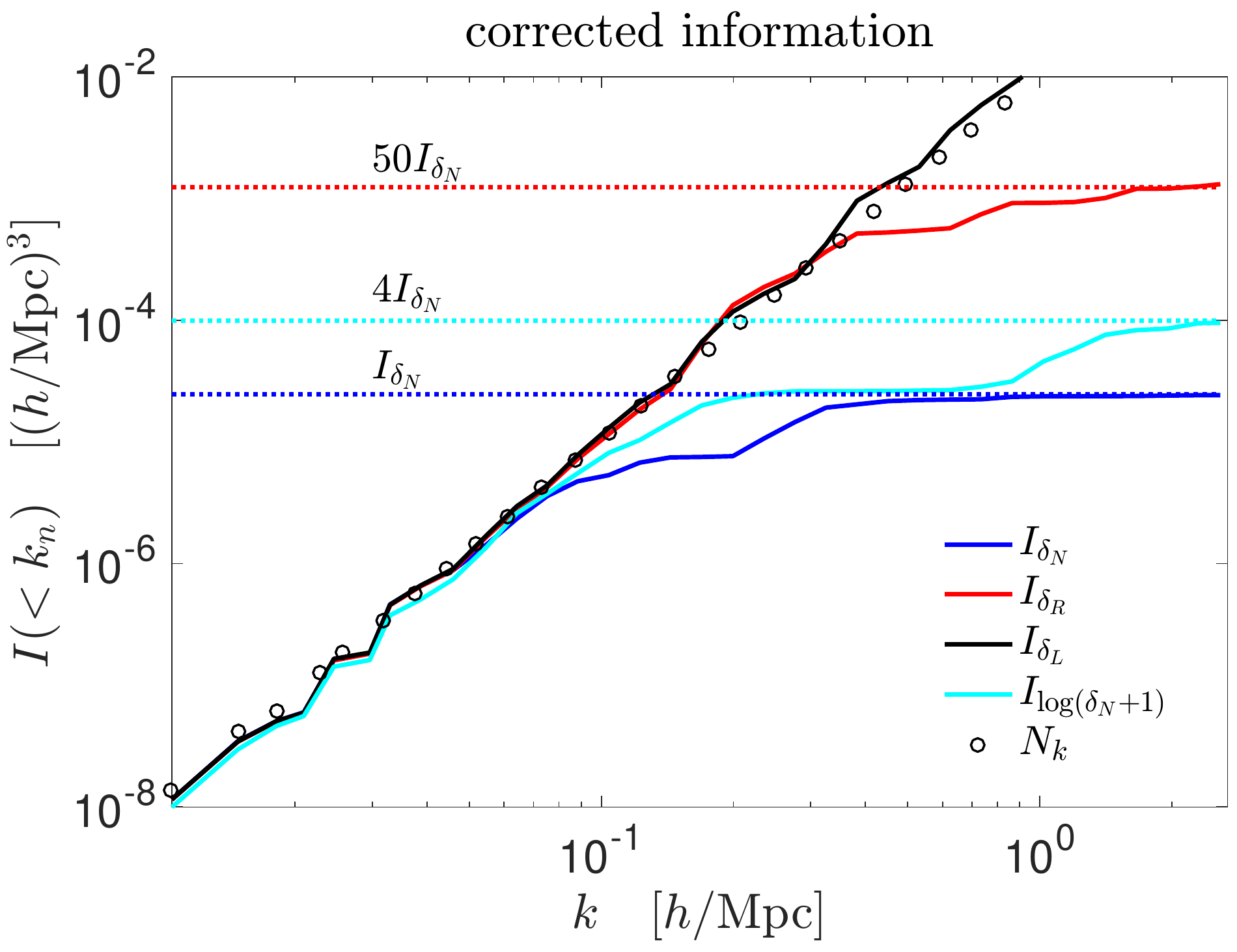}
    \includegraphics[width=0.48\textwidth]{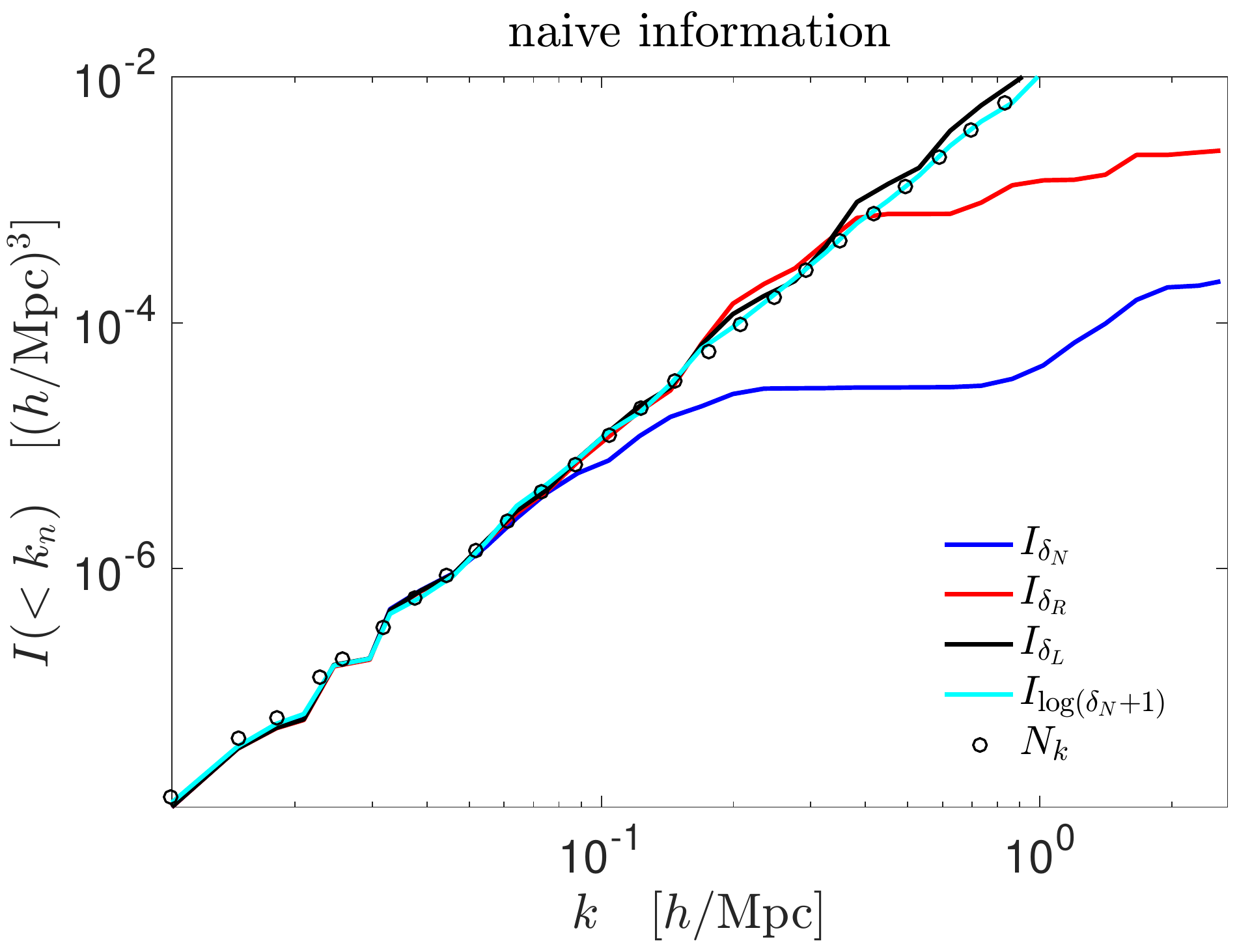}
    \centering
    \caption{{\it Left.} The Fisher information (solid lines) per unit volume as
      a function of wave number.  The blue, red, and black curves correspond to the power spectra
      of $\delta_N$, $\delta_R$ and $\delta_L$ respectively,
      and the cyan curve
      corresponds to the logarithmic density mapping. The circles
      are the cumulative number of $k$ modes.  Dotted horizontal lines indicate the value of the 
      Fisher information at $k \simeq 2.7$ $h$/Mpc.  {\it Right.} Same
      as the left panel except setting $r\equiv 1$ in Eq.~\ref{eq:fisherformulaused}. The
      black, blue, and cyan lines match the results in \cite{bib:Rimes2006,bib:Mark2009}. 
This naive estimate does not account for the lack of correlation
of the final density field with the initial conditions, and represents
an overestimate of the information.}
  \label{fig:fisherinfo}
\end{figure*}
\end{section}

\begin{section}{Conclusion}
  \label{sec:conclusion}
  We study the Moving-Mesh algorithm's ability to estimate the underlying displacement 
  field and reconstruct the linear density fields using 130 cosmological $N$-body 
  simulations.  We measure the power spectra and the associated covariance of the 
  nonlinear density fields, and the reconstructed density fields.  We quantify the 
  result by (i) cross-correlating them with the linear density fields, (ii) studying 
  the $k$-mode coupling in the correlation matrix, and (iii) computing the cumulative 
  Fisher information contained in these power spectra.  We also
  compared with the $E$-mode 
  reconstruction and logarithmic density mapping Gaussianization
  techniques.
We find that Moving-Mesh method gives better results than previous works
  (e.g. \cite{bib:Mark2009,bib:Zhang2011,bib:HarnoisD2013}), and on scales 
  relevant to the BAO, our result approaches the optimal $E$-mode reconstruction.  
  Future steps include quantifying Halo Poisson noise and bias contamination 
  from realistic measurements, and determining the quantitative impact on
  BAO and RSD measurements.  

%
\end{section}

\acknowledgements{
  \label{sec:acknowledgements}
  We thank Xin Wang and Kwan Chuen Chan for friendly and helpful discussions.  We thank 
  Hong-Ming Zhu for discussions and detailed comments and feedback on drafts of this work.
  We thank Yu Yu for discussions and providing an optimized version of the reconstruction code. 
  Computations were performed on
  the General Purpose Cluster supercomputer at the SciNet HPC
  Consortium.  SciNet is funded by: the Canadian Foundation for
  Innovation under the auspices of Compute Canada; the Government of
  Ontario; Ontario Research Fund - Research Excellence; and the
  University of Toronto.  
}

\bibliographystyle{apsrev}
\bibliography{myreference}

\end{document}